\documentstyle[prl,aps,tighten,multicol,epsf]{revtex}

\begin{document}

\draft

\title{Thermodynamical fingerprints of fractal spectra}

\author{Ra\'ul O. Vallejos$^1$ and 
       Celia Anteneodo$^2$}

\address{$^1$
	 Instituto de F\'{\i}sica, 
	 Universidade do Estado do Rio de Janeiro, \\
	 R. S\~ao Francisco Xavier, 524, CEP 20559-900 Rio de Janeiro, 
	 Brazil \\
         {\rm e-mail: vallejos@cat.cbpf.br} \\
	 $^2$
	 Instituto de Biof\'{\i}sica, Universidade Federal do Rio de
         Janeiro, \\ 
         Cidade Universit\'aria, CCS Bloco G, CEP 21941-900,\\
         Rio de Janeiro, Brazil \\
         {\rm e-mail: celia@cat.cbpf.br}}

\date{\today}

\maketitle

\begin{abstract}
We investigate the thermodynamics of model systems exhibiting two-scale
fractal spectra. In particular, we present both analytical and
numerical studies on the temperature dependence of the vibrational and
electronic specific heats.  For phonons, and for bosons in general, we
show that the average specific heat can be associated to the average
(power law) density of states.  The corrections to this average
behavior are log-periodic oscillations which can be traced back to the
self similarity of the spectral staircase. In the electronic case, even
if the thermodynamical quantities exhibit a strong dependence on the
particle number, regularities arise when special particle numbers are
considered. Applications to substitutional and hierarchical structures
are discussed.
\end{abstract}


\begin{multicols}{2}

\narrowtext


\section{Introduction}


In the last 15 years quasiperiodic structures have been studied
intensively. Aside from their purely theoretical interest, these
studies were in part motivated by the discovery of the quasicrystalline
state\cite{schechtman84} together with the possibility of experimental
realization of quasiperiodic superlattices, first achieved in 1985 by
Merlin and collaborators\cite{merlin85}. In turn, the rich properties
of these structures encouraged the study of systems based on
alternative sequences (e.g., Thue-Morse and hierarchical), which, in
spite of not being quasiperiodic, still exhibit deterministic (or
``controlled'') disorder, i.e., they are nor random neither periodic.
One of the most interesting features that many of these problems
(either experimental or theoretical) display is a fractal spectrum of
excitations.  For instance, experiments on Thue-Morse\cite{merlin88}
and Fibonacci\cite{merlin85} superlattices have uncovered scale
invariant energy spectra.  Numerical analysis of linear chains of
harmonic oscillators with hierarchical nearest-neighbor couplings and
equal masses exhibit spectra related to the triadic
Cantor-set\cite{petri95} (the Cantor-like structure is preserved even
if masses are also distributed in a hierarchical way\cite{kimball97}).
It has also been proved that the energy spectrum of a chain made of
identical springs and of masses of two different kinds arranged after
the Thue-Morse sequence is a Cantor-like set\cite{axel89}.

Even though controlled disorder typically leads to multifractal spectra,
in some cases only a few scales are sufficient for a satisfactory
understanding of their thermodynamics. For instance, the phonon
spectrum of the chain in \cite{petri95} is essentially a one-scale
Cantor-set; the electronic spectra that arise from Fibonacci
tight-binding Hamiltonians (either on-site or transfer) are governed by
a couple of scale-factors\cite{niu86}.

Within this context, few-scale fractal spectra constitute simple
prototypes for testing the thermodynamical implications of
deterministic disorder. As a first step in this direction, in
Refs.~\cite{tsallis97,vallejos98}, one- and multi-scale fractal energy
spectra were studied within Boltzmann statistics. It was shown that the
scale invariance of the spectrum has strong consequences on the
thermodynamical quantities. In particular, the specific heat oscillates
log-periodically as a function of the temperature. Moreover, general
scaling arguments and a detailed analysis of the integrated density of
states allowed for a quantitative prediction of the average value
(which is related to the average density of states), period and
amplitude of the oscillations.

In this paper we extend the analysis of \cite{tsallis97,vallejos98} to
$N$-particle systems described by {\em quantum} statistics. We present
results concerning the relevant thermodynamical quantities, i.e.,
chemical potential, average particle number and specific heat. Our
conclusion is that for phonons, and for bosons in general, the
Boltzmann scenario survives the inclusion of quantum symmetries. That
is, the crudest (power-law) approximation to the density of states
leads to a correct prediction of the average behavior of the specific
heat. Log-periodic oscillations decorating the average value are well
reproduced when the simplest non-trivial corrections to the power-law
density of states are considered. In the electronic case, the presence
of a Fermi surface makes the thermodynamical quantities very sensitive
to the particle number, thus preventing the use of smooth
approximations.  However, for special cases, interesting regularities
can still be observed.

The paper is organized as follows. In Section II we discuss some
general aspects of the problem. We make as well a short revision of
previous results which will be useful in subsequent sections.  In
particular, we introduce the spectra that will be considered and
describe their significant characteristics.  In Section III we treat
the phonon case, whose essential features can be explained analytically
by means of scaling arguments similarly to the Boltzmann case. Section
IV is devoted to Bose (general case) and Fermi statistics. Section V
contains the concluding remarks.


\section{General considerations}


We consider a spectrum constituted by levels lying on a $(r_1,r_2)$
fractal set\cite{tel88}. Such spectrum can be constructed recursively
starting from an arbitrary discrete set of levels $\{\epsilon^{(0)}\}$
belonging to the interval $[0,\Delta]$.  The hierarchy
$\{\epsilon^{(n+1)}\}$ consists of two pieces which are obtained by
compressing the set $\{\epsilon^{(n)}\}$ by factors $r_1$ and $r_2$,
respectively. The lowest end of the first piece is positioned at
$\epsilon=0$, the upper end of the second piece is positioned at
$\Delta$ (see Fig.~1). The fractal that arises in the limit
$n\rightarrow \infty$ has a (Hausdorff) dimension $d_f$ given
implicitly by $r_1^{d_f}+r_2^{d_f}=1$. In the particular one-scale
case, i.e., $r_1=r_2\equiv r$, one has the explicit result $d_f=-\ln
2/\ln r$.  An alternative way of viewing the construction of the
fractal is that in terms of gaps and ``bands''.  At hierarchy $n$ there
are $2^n-1$ gaps (in gray in Fig.~1), which define $2^n$
``bands'' (in white). When $n$ is increased by one, the preexistent
gaps remain untouched, but each band splits into two new ones and a new
gap appears.

We expect the thermodynamical properties associated to this spectrum to
depend only on its hierarchical organization and not on the specific
pattern for $n=0$\cite{vallejos98}. For this reason, and for the sake of
notation simplicity, we will take $\{\epsilon^{(0)}\}=\{0,\Delta\}$ (so
that there are $2^{n+1}$ levels at generation $n$).  In addition, we
make $\Delta=1$ and $k_B=1$ ($k_B$ stands for Boltzmann's
constant) throughout the paper.
\vspace*{1.0pc} 
\begin{figure}[ht]
\hspace{1.0pc} 
\epsfxsize=6.5cm
\epsfbox[109 327 437 477]{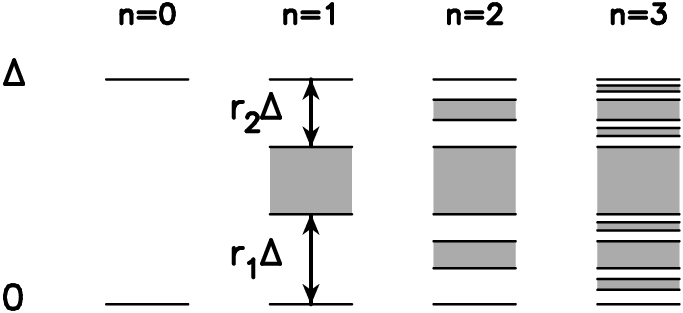}
\vspace*{1.0pc} 
\caption{The two-scale energy spectrum.  Shown are the first
hierarchies in the construction of a $(r_1,r_2)$ fractal spectrum. The
initial pattern ($n=0$) was chosen as a set of two levels,
$\{\epsilon_1^{(0)},\epsilon_2^{(0)}\}=\{0,\Delta\}$. At generation
$n$, $2^{n-1}$ new gaps (in gray) appear.  The (positive) scale factors
are restricted by the condition $r_1+r_2 \le 1$.}
\end{figure}
A very useful tool in describing Boltzmann systems
with fractal spectra\cite{tsallis97,vallejos98} is the integrated
density of states ${\cal N}(\epsilon)$ defined as
\begin{equation} 
{\cal N}(\epsilon)
\equiv\int_0^\epsilon {\rm d}\epsilon' D(\epsilon') ~,  
\end{equation}
where $D(\epsilon)$ is the one-particle state density. In spite of the
very fragmented structure of the density of states (a set of delta
functions with fractal support), a smooth approximation is usually good
enough for computing integral quantities, e.g., the specific heat. In
the present case, the scaling properties of the spectrum imply that in
a log-log plot ${\cal N(\epsilon)}$ looks as a self-similar staircase
with a fractional average slope. If first-order log-periodic
corrections are included, we obtain a smooth approximation of the
form\cite{vallejos98}
\begin{equation}  \label{smooth}
{\cal N}(\epsilon)/2^{n+1} \approx \epsilon^d
[a\,+\,b\cos(\omega\ln\epsilon-\phi)] ~.  
\end{equation}
The so-called spectral dimension $d$\cite{rammal83} and the 
frequency $\omega$
arise from simple scaling arguments, yielding
\begin{equation}  \label{dimomega}
d=-\ln 2/\ln r_1 ~~;~~ \omega=-2\pi/\ln r_1 ~.  
\end{equation}
The parameters $a \approx 1$, $b$, and $\phi$ are functions of both
$r_1$ and $r_2$ and result from a detailed analysis of the exact
spectral staircase. The smooth cummulative density (\ref{smooth})
allows for analytical manipulations while still keeping the principal
ingredients of the self-similar spectrum. In fact, in the Boltzmann
case, the use of (\ref{smooth}) together with a perturbative approach
leads to an accurate expression for the specific heat per particle as a
function of the temperature $T$:
\begin{equation}  \label{CMB}
c(T) \approx d + a'\cos(\omega\ln T-\phi) ~,
\end{equation}
where $a' \ll d$ is a function of $r_1$ and $r_2$\cite{vallejos98}. 
Eq.~(\ref{CMB}) relates explicitly the log-periodic nature of the 
specific heat to that of the spectral staircase.


\section{Phonon statistics}


A standard way to study the acoustic properties of a lattice is to
consider a nearest-neighbor harmonic chain. This model is represented
by the equation of motion
\begin{eqnarray} \label{chain} 
m_j\frac{d^2u_j}{dt^2} & = &     
k_{j,j+1} u_{j+1} +   k_{j,j-1} u_{j-1} \nonumber \\
                       &   &       
- ( k_{j,j+1} + k_{j,j-1} ) u_{j}   ~,
\end{eqnarray}
where $u_j$ is the displacement of the $j$th atom (of mass $m_j$) from
its equilibrium position. $k_{j,j\pm1}$ are the strengths of the
couplings between neighboring atoms. Deterministic disorder is
introduced by requiring the masses and/or the strengths to follow
substitutional (e.g., Thue-Morse, Fibonacci,
Fibonacci-class\cite{fu97}) or hierarchical sequences. Assuming that
the time dependence in (\ref{chain}) goes like $u_j \propto
\exp(-i\omega t)$, the stationary equation of motion is obtained. Upon
diagonalization one calculates the normal modes and the
eigenfrequencies $\omega$. In the case of Thue-Morse\cite{axel89},
Fibonacci\cite{luck86,nori86,lu86,liu87} and
Fibonacci-class\cite{anteneodo98} sequences the spectrum of energies
$\epsilon=\hbar\omega$ has a Cantor-like structure only in the {\em
short} wavelength regime. For long wavelengths the number of gaps and
their sizes tend to zero and one recovers the standard results for the
periodic lattices.  On the other side, the hierarchical chains of
Refs.~\cite{petri95,kimball97} do possess spectra with uniform scaling
of the type ($r_1,r_2$). For these cases we can apply the formalism of
\cite{vallejos98} to obtain an explicit expression for the specific
heat. The specific heat per atom $c(T)$ is calculated as
\begin{equation} \label{CP}
c(T)=\frac{1}{2^{n+1}} 
\sum_{j=1}^{2^{n+1}} \left[  
\frac{\epsilon_j/2T}{\sinh(\epsilon_j/2T)} \right]^2 ~,
\end{equation}
where the summation runs over all energy levels of a ($r_1,r_2$)
fractal of $n$-th generation.  In Fig.~2 we illustrate the dependence
of the specific heat (\ref{CP}) with temperature for different values
of $(r_1,r_2)$.  Logarithmic scales have been used to
clearly display that, in the low temperature regime ($T \ll 1$), a)
$c(T)$ increases in average as power law; b) there are log-periodic
corrections to the average power law. As $T$ is lowered this behavior
persists down to a minimum temperature $T_{min} \approx r_1^n$,
corresponding to the smallest scale of the fractal truncated at
hierarchy $n$ (when $n \rightarrow \infty$, $T_{min} \rightarrow 0$).
Of course, for high temperatures ($T \gg 1$), the classical result
$c(T)=1$ for a one dimensional lattice is recovered.  Features a) and
b) can be shown to be a direct consequence of the log-periodicity of the
spectral staircase, which is captured by the smooth approximation
(\ref{smooth}). In order to show that this is the case let us first
rewrite (\ref{CP}) as an integral:
\begin{equation} \label{CPint}
c(T)=
\frac{1}{2^{n+1}} \int_0^1
\left[ \frac{\epsilon/2T}{\sinh(\epsilon/2T)} \right]^2
d{\cal N}(\epsilon)  ~.
\end{equation}
Now we make two approximations in Eq.~(\ref{CPint}). First we replace
the exact ${\cal N}(\epsilon)$ by its smooth version (\ref{smooth}).
Second: as our interest is the low-$T$ regime, we extend the upper
integration limit to infinity (this is equivalent to considering an
unbounded fractal spectrum). After some simple algebra we arrive at
\begin{eqnarray} \label{CPx}
c(T) & \approx & T^d \, d \, 
                  [ a'' + 
                    b'' \cos(\omega\ln T-\phi) + \nonumber \\
     &         &    c'' \sin(\omega\ln T-\phi) ] ~.
\end{eqnarray}
Here 
$a''=aa'$, $b''= b\,(b'-c'\omega /d)$, and $c''= -b\,(c'+b'\omega /d)$; 
the parameters $a'$, $b'$, $c'$ are given by 
\begin{equation} \label{abc}
\left[ 
\matrix{ a'\cr b'\cr c' \cr } 
\right] 
= \frac{1}{4} \int_0^{\infty} dx \, \frac{x^{d+1}}{\sinh^2(x/2)}
\left[ 
\matrix{ 1\cr \cos(w\ln x)\cr \sin(w\ln x) \cr} 
\right] . 
\end{equation} 
Eq.~(\ref{CPx}) is the phonon analogue of the Boltzmann result
(\ref{CMB}). It displays explicitly the average behavior of the
specific heat and the log-periodic corrections. In Fig.~2 we have
plotted the analytical curves (\ref{CPx}) for the same set of scaling
factors used for the numerically exact calculations.
\begin{figure}[ht]
\hspace{0.0pc} 
\epsfxsize=7.0cm
\epsfbox[66 237 482 555]{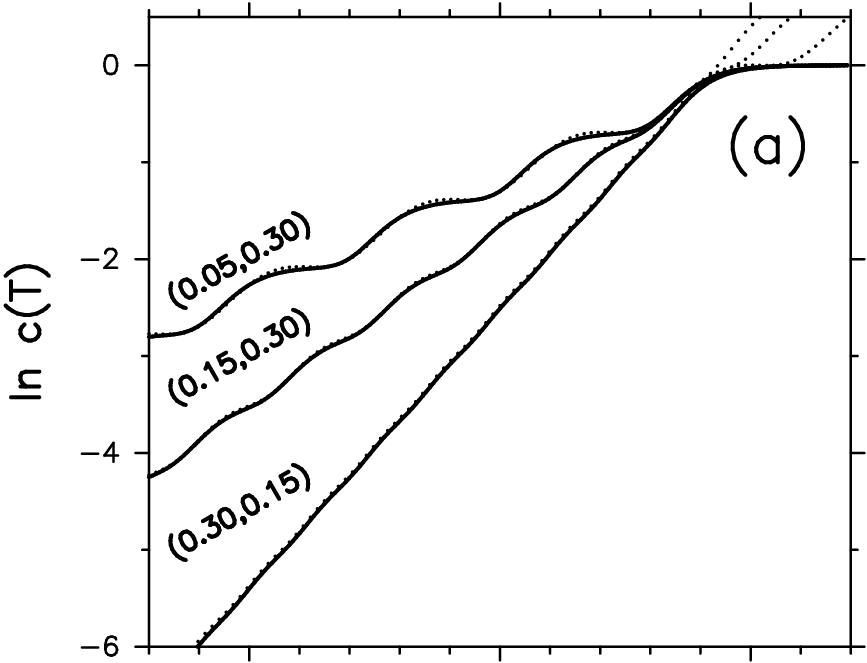}
\end{figure}
\vspace*{-1.5pc} 
\begin{figure}[ht]
\hspace{0.0pc} 
\epsfxsize=7.0cm
\epsfbox[66 237 482 555]{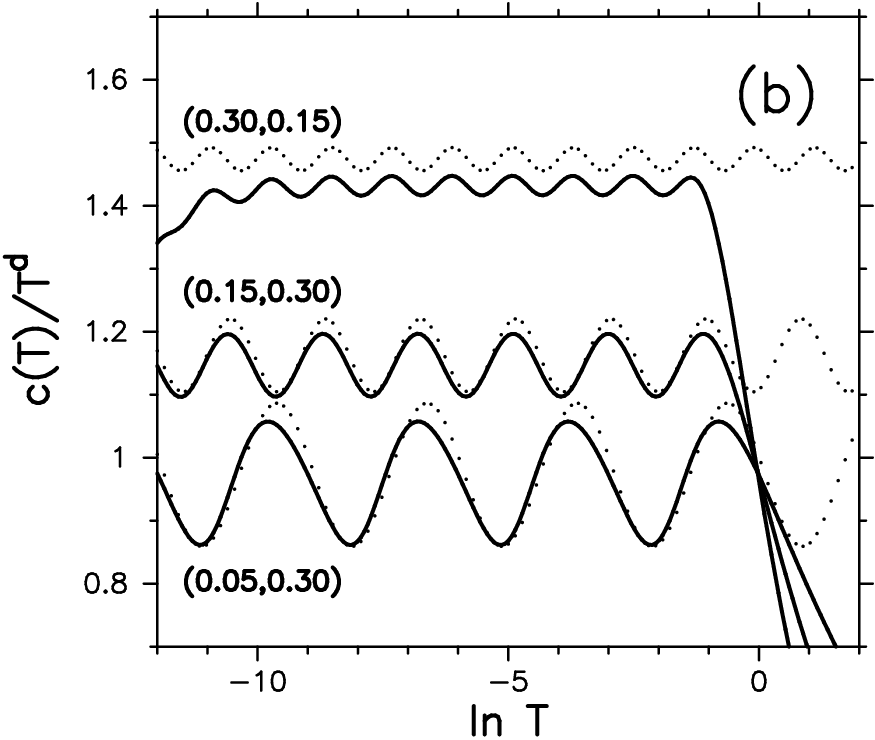}
\vspace*{1.0cm} 
\caption{Phonons in a two-scale fractal spectrum. (a) Log-log plot of
the specific heat per atom versus temperature. The curves are
parametrized by the scale factors $(r_1,r_2)$ indicated in the figure.
(b) $c(T)/T^d$ vs $\ln T$.  In all cases $n=9$. Full lines correspond
to numerical results (\ref{CP}) and dotted ones to our analytical
prediction (\ref{CPx}), which is valid only for low temperatures (this
is the reason why the analytical curves do not saturate at high
temperatures).}
\end{figure}
For low temperatures, the agreement between the exact and analytical
results is very good [notice that the disagreement in the  mean value
for the $(0.30,0.15)$ case actually corresponds to a $3\%$ error]. The
scale factor $r_1$ is responsible for the period of the oscillations
through $\omega=-2\pi/\ln r_1$ (as shown in Fig.~2b the period
decreases as $r_1$ increases).  Moreover, as $r_1$ controls the
gap-scaling with respect to $T=0$, smaller $r_1$ values imply
comparatively bigger gaps, and, in turn, bigger amplitudes.  
For fixed $T$, decreasing $r_1$ causes the spectrum to shrink 
towards $\epsilon=0$, as a consequence the specific heat grows. 
These are the features observed by Petri and Ruocco in their
paper on one dimensional chains with hierarchical
couplings\cite{petri95}. We can now understand quantitatively 
their results as being a direct consequence of log-periodic corrections 
to the pure power-law scaling of the density of states (\ref{smooth}).


\section{Bose and Fermi statistics}


The statistical problem of particles with non identically null chemical
potential stands at the next level of complexity, both from analytical
and numerical points of views. As in this case analytical derivations
seem not feasible (except in limiting regimes), we resort to a
numerical procedure. For a fixed {\em average} number of particles $N$
we extracted the chemical potential $\mu$ as a function of the
temperature by solving the equation
\begin{equation}  \label{NBF}
N = \sum_{j=1}^{2^{n+1}}  \frac{1}{e^{(\epsilon_j-\mu)/T} \pm 1} ~, 
\end{equation}
where the sum runs over the levels of the fractal truncated at
hierarchical depth $n$. The sign $+/-$ corresponds to the Fermi/Bose
case. (We are not taking into account spin degeneracies, which will not
modify our results.) The numerical solution of Eq.~(\ref{NBF}) is not a
difficult task, however some care has to be taken with the choice of
an appropriate initial value for $\mu$.  Once the chemical potential
has been obtained, the specific heat can be calculated as the
$T$-derivative of the {\em average} total energy $E(T)$, 
\begin{equation} \label{EBF}
E(T)= \sum_{j=1}^{2^{n+1}}  
      \frac{\epsilon_j}{e^{(\epsilon_j-\mu)/T} \pm 1} ~. 
\end{equation}
%


\subsection{Bosons}


Even though bosonic excitations with non-null chemical potential do not
seem to be of main relevance for the thermodynamics of
superlattices, we decided to include a discussion on this topic for
reasons of completeness, and because bosons in a fractal spectrum
display a paradigmatic behavior.

The dependence of the fugacity $z=\exp(\mu/T)$ with temperature is
illustrated in Fig.~3a. Several curves, parametrized by the average
particle number (indicated in the figure) have been drawn.  First of
all, Fig.~3a displays the following limiting features.  As $T$
decreases, and the system evolves towards condensation, the chemical
potential increases from negative values up to a limiting value close
to zero (we recall that our lowest energy is $\epsilon=0$). The number
of particles in the condensate is given by $N_0=z/(1-z)$.  Thus, for
very low $T$, together with large $N$, one has $z \approx 1-1/N$. In
the opposite limit of high temperatures and small $N$, one obtains from
(\ref{NBF}) $z\rightarrow N^* \equiv N/2^{n+1}$. In Fig.~3b we have
chosen a small value of $r_1$ to show that $\mu$ 
indeed oscillates as the temperature gets through the scales of the
spectrum. In this sense, Fig.~3b can be thought as an
amplified version of Fig.~3a.
\begin{figure}[ht]
\hspace{0.0pc} 
\epsfxsize=7.0cm
 \epsfbox[79 237 482 555]{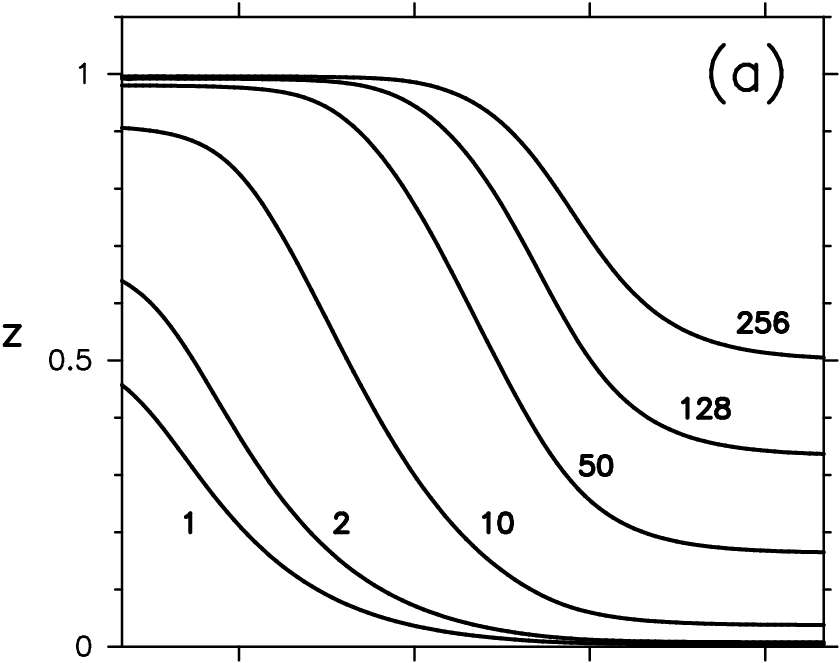}
\end{figure}
\vspace*{-2.0pc}
\begin{figure}[ht]
\hspace{0.0pc} 
\epsfxsize=7.0cm
 \epsfbox[79 200 482 555]{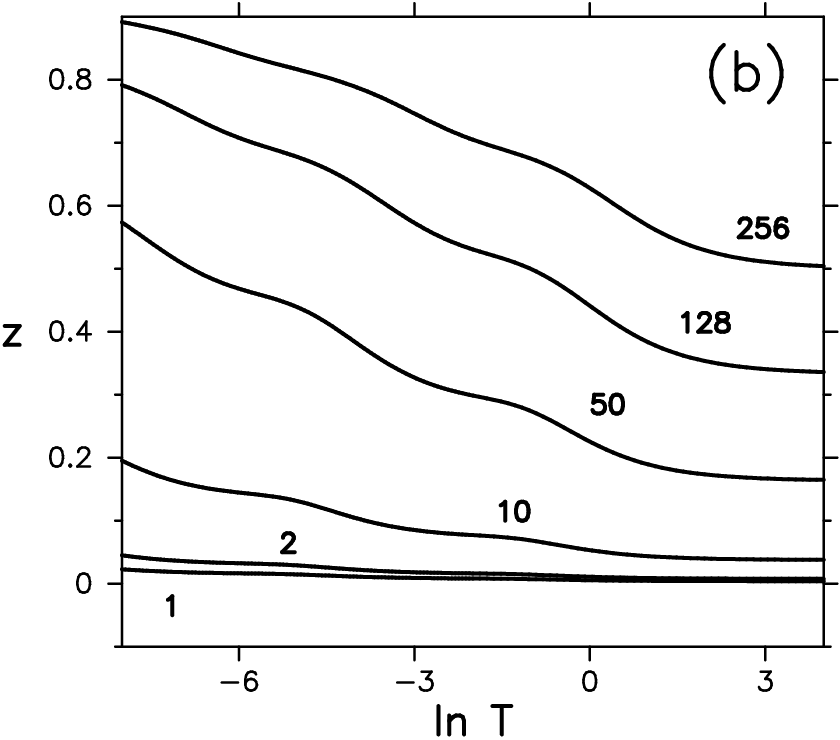}
\vspace*{1.0pc} 
\caption{Bosons in a two-scale fractal spectrum.  Fugacity $z$ vs. $\ln
T$, for different values of particle number $N$ (indicated in the
figure).  The chosen scales are $(r_1,r_2)=(1/3,0.2)$ in (a) and
$(0.02,0.2)$ in (b). In both cases $n=7$.}
\end{figure}
The dependence of the specific heat per particle on the temperature is
illustrated in Fig.~4. Shown are the exact result and the numerical
calculation which uses the smooth expression (\ref{smooth}) for the
spectral staircase. For low temperatures the agreement is excellent.
It is clear that the mean value of the specific heat is indeed
associated to the average level density and that it suffices to take
into account only the first non-trivial correction to the power-law
scaling to explain the oscillations. [It is worth pointing out that
this scenario breaks down in the fermionic case (see later)].  We have
tested that the curves in Fig.~4 do not change in the thermodynamical
limit of increasing both $N$ and the hierarchy $n$ but keeping
$N^*=N/2^{n+1}$ fixed, except that the oscillatory behavior extends to
lower temperatures of the order of $r_1^n$ (the total number of
oscillations of the specific heat is equal to the depth $n$).  The
relative particle number $N^*$ plays the role of a density, given
that the number of levels below a certain energy grows with the volume
of the system. When $N^*$ is sufficiently small and the temperature
is high enough, the curves tend to those for the Boltzmann
statistics\cite{tsallis97,vallejos98}. This is analogous to the usual
statement that the parameter $\lambda^3 \rho$ (where $\lambda$ and
$\rho$ are the thermal length and the density, respectively) tells
how good is the classical approximation.
\begin{figure}[ht]
\hspace{0.0pc} 
\epsfxsize=7.0cm
 \epsfbox[66 193 482 555]{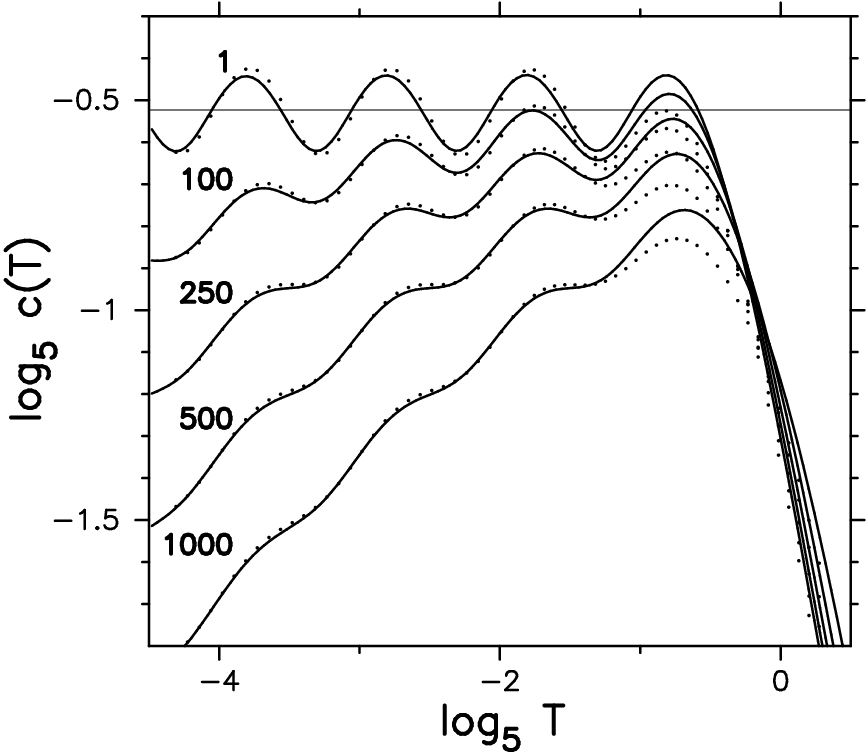}
\vspace*{1.0pc} 
\caption{Bosons in a two-scale fractal spectrum.  Log-log plot of the
exact specific heat per particle $c(T)$ vs. $T$, for different values
of average particle number $N$ (full lines).  Dotted lines represent a
calculation which uses the cummulative density of states
(\ref{smooth}).  The horizontal line is $c(T)=d$, which is the
classical average value for low $T$.  At high temperature, the specific
heat decays according to the classical $1/T^2$ law. Scale factors are
$(r_1,r_2)=(1/5,1/3)$. The fractal has depth $n=9$. Base-5 logarithms
were used to show that there is one oscillation per ``decade''.}
\end{figure}

Although an analytical description of the problem is lacking, our numerical 
calculations show that log-periodicity is robust enough to resist the 
inclusion of bosonic symmetries together with the restriction of particle 
conservation.

\subsection{Fermions}


The results presented in this section are valid for fermions in
general, however, we assume that the fractal $(r_1,r_2)$ corresponds to the
electronic spectrum of a certain superlattice. From the theoretical
point of view, such spectra appear in the simplest model for studying
the electronic properties of a lattice, i.e., a stationary
tight-binding equation, for instance, in its transfer version:
\begin{equation}
t_{j+1} \psi_{j+1} + t_j \psi_{j-1} = \epsilon \psi_j ~,
\end{equation}
where $\psi_j$ denotes the wave function at site $j$ and $\{t_j\}$ are
the hopping matrix elements. If these hopping elements are arranged
according to the Fibonacci sequence, the spectrum of energies
$\{\epsilon\}$ is essentially a fractal of the type $(r_1,r_2)$ (see
e.g. \cite{niu86}), independently of the boundary conditions.  However,
the simple scaling is lost if the more general
Fibonacci-class\cite{fu97} sequences are considered.

The numerical scheme used for computing the thermodynamical quantities
in the bosonic case can be easily adapted to the electronic problem.
Figs.~5 and 6 exhibit some relations among chemical potential $\mu$,
average number of particles $N$ and temperature for the $r_1=r_2=1/3$
case.  Fingerprints of the fractal spectrum are clearly seen in the
dependence of $\mu$ with temperature (Fig.~5). In the limit of zero
temperature, for a given integer particle number $N$, $\mu$ takes the
value corresponding to the middle of the gap
$(\epsilon_N,\epsilon_{N+1})$, as happens, e.g., in intrinsic
semiconductors. In order to keep $N$ fixed as temperature grows, the
Fermi surface moves in the direction of the lowest density of levels,
integrated over an energy interval $T$. The relative concentration of
levels in the neighborhood of the gap is a fluctuating function of the
scale $k_BT$.  This gives rise to an oscillatory process that persists
until $T$ overcomes the largest gap ($\log_3 T\approx -1$).
From this point on, if $N^*>0.5$ ($N^*<0.5$), $\mu$ tends
in a monotonic way to $+\infty$ ($-\infty$).
\begin{figure}[ht]
\hspace{0.0pc} 
\epsfxsize=7.0cm
\epsfbox[74 193 482 555]{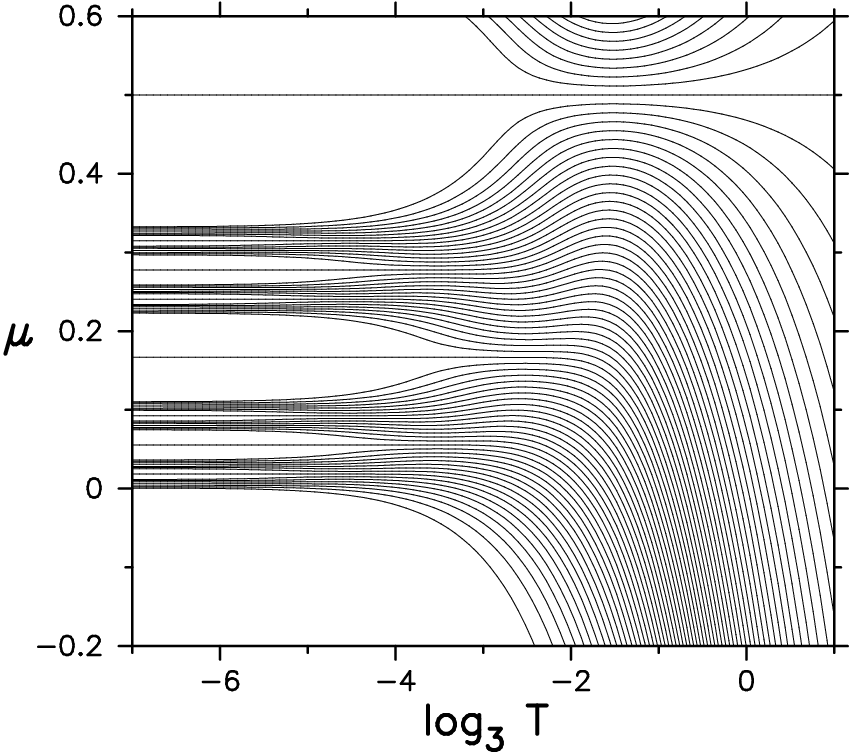}
\vspace*{1.0pc} 
\caption{Fermions in a fractal spectrum. Chemical potential $\mu$ as a
function of $\log_3 T$ for all integer particle numbers $1 \le N \le
2^{n+1}=128$ (increasing from bottom to top). We have only plotted the
lower part of the full figure because it is symmetric with respect to
$\mu=0.5$.  Parameters are $r_1=r_2=1/3$, $n=6$.}
\end{figure}
For smaller values of $r_1$, oscillations in $\mu$ are magnified (not
shown), analogously to what happens with bosons.

Fig.~6 displays the chemical potential as a function of the reduced
number of particles $N^*$ for three values of fixed temperature. 
This figure gives some insight of the difficulties involved in solving
Eq.~(\ref{NBF}). For low (fixed) temperatures, $\mu(N)$ tends to the
spectral staircase and the numerical problem gets relatively difficult.
As temperature grows the staircase gets smoother and numerics simplify. 
We have taken $N$ to be continuous, but observe that for integer $N$
and low temperatures (compared to the gap size) the curves pass through 
the middle of the gaps, as it is clear in Fig.~5.  
\begin{figure}[ht]
\hspace{0.0pc} 
\epsfxsize=7.0cm
\epsfbox[79 193 482 555]{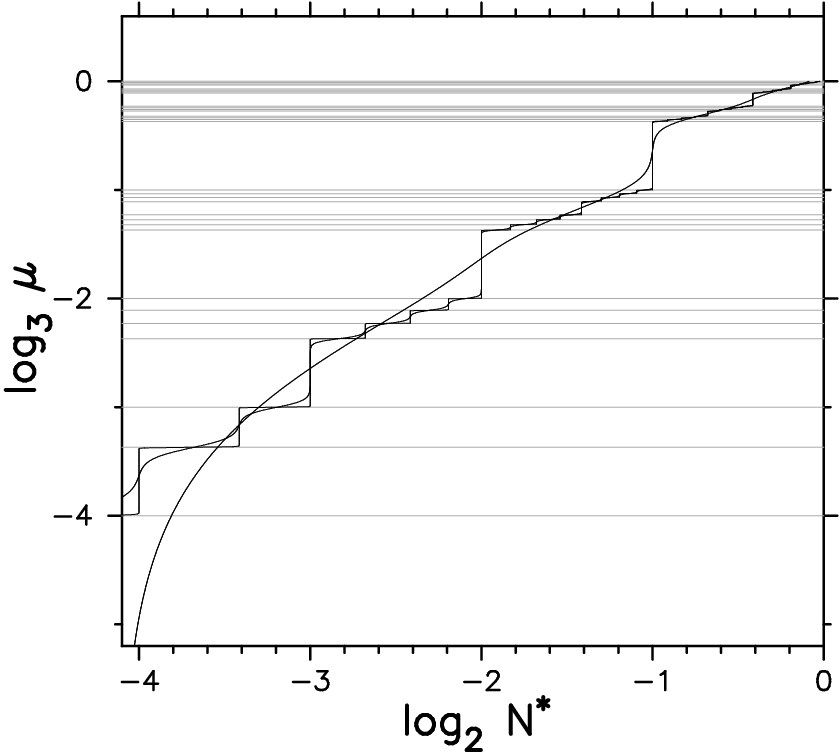}
\vspace*{1.0pc}
\caption{Fermions in a fractal spectrum. Log-log plot of the chemical
potential $\mu$ versus the reduced number of particles $N^*$ for
fixed temperature.  The curves correspond to $T=3^{-9}$, $3^{-6}$, and
$3^{-3}$ (smoother curves correspond to higher temperature).  Gray
lines indicate the energy levels. We used $r_1=r_2=1/3$ and $n=4$.}
\end{figure}
The specific heat in the fermionic problem depends strongly on the 
position of the Fermi surface. Thus, we begin by analyzing some 
simple cases where the number of particles takes special values. 
For this purpose it is useful to recall the picture of bands and gaps 
of Fig.~1. 
Each gap determines a special number of particles $N^*$, namely, 
the relative number of particles that would lie below that gap at 
$T=0$. For instance, the first-generation gap determines $N^*=1/2$. 
The $n=2$ gaps correspond with $N^*=1/4,3/4$. 
Conversely, the denominator and the numerator in $N^*$
indicate respectively the generation $n$ of a gap and the number of
filled bands of $n$th generation, e.g., $N^*=1/32=1/2^5$ corresponds to
a gap which {\em appeared} in the fifth generation and to one filled band
of fifth generation.  

A first example that illustrates the structure of the specific heat is
depicted in Fig.~7. There we consider a spectrum with two different
scales $r_1=1/3$, $r_2=1/9$, and the selected set of numbers
$N^*=1/32, 1/16, 15/16, 31/32$.  For the first two cases,
$N^*=1/32,1/16$, the size of the gap is of the order of the Fermi
temperature $T_F$. Once the temperature overcomes the width of that
gap, the electrons can fully access the next band above, so that the
mean level occupation drops approximately to 1/2.
From then on the electrons behave essentially as Boltzmann particles:
the specific heat oscillates around a constant average value, as
expressed by Eq.~(\ref{CMB}). The average value and the period are
governed by the scaling factor $r_1=1/3$ through Eq.~(\ref{dimomega}).
The number of oscillations is equal to the generation index $n$:  four if
$N^*=1/16$ and five if $N^*=1/32$. The case of numbers $N^*=31/32,15/16$
is equivalent to having respectively $N^*=1/32,1/16$ {\em holes} and it
is then complementary to the previous one. Remarkably, in the case of
holes, the dominant scale is $r_2$.  The reason for this is that $r_2$
rules the scaling with respect to the upper edge of the spectrum, and,
in fact, with respect to the upper (lower) edge of every band (gap); in
other words, $r_2$ is the relevant factor for {\em negative}
temperatures. Eqs.~(\ref{dimomega}) and (\ref{CMB}) are still valid
provided that $r_1$ and $r_2$ are interchanged. According to these
arguments, the choice $r_2=r_1^2$ implies that the period of the
oscillations for holes is twice that for electrons, which is confirmed
by our calculations (see Fig.~7). Horizontal lines correspond to the
Boltzmann average result, properly normalized in the case of holes.
Increasing the hierarchy $n$ and keeping $N^*$ fixed does not change
Fig.~7 because the gap on top of the filled band determines the
smallest scale of the fractal that can be resolved.
\begin{figure}[ht]
\hspace{0.0pc} 
\epsfxsize=7.0cm
\epsfbox[77 193 482 555]{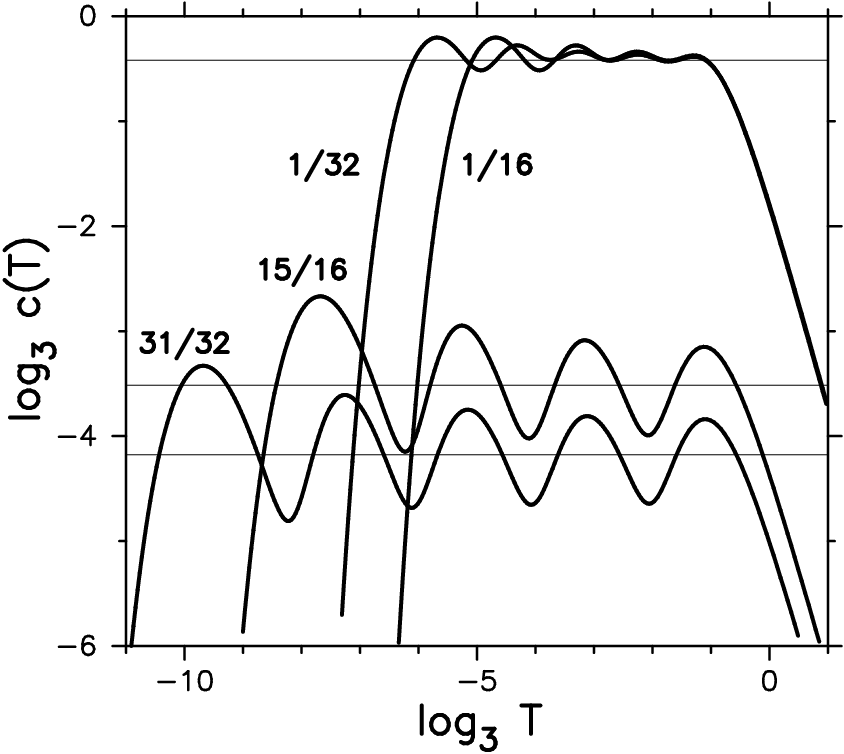}
\vspace*{1.0pc}
\caption{Fermions in a fractal spectrum. Log-log plot of the specific
heat versus temperature. The reduced numbers of particles are
$N^*=1/32,1/16,15/16,31/32$, and the scale factors $r_1=1/3$,
$r_2=1/9$.  The horizontal lines represent the classical average value
for the appropriate number of ``hot'' electrons ($N^*=1/32,1/16$) or
holes ($N^*=15/16,31/32$).}
\end{figure}
Other special particle numbers allow for observing a
mixed electron-hole behavior. Let us consider, for instance, the cases
$N^*=7/64,57/64$, corresponding, respectively, to 7 and 57 filled bands
of sixth generation (Fig.~8).  For $N^*=7/64$, the Fermi temperature is
several scales bigger than the width of the seventh gap, so
that the low-$T$ part of the specific heat is associated to hole
excitations jumping from the (empty) eighth band down over the sixth-,
fifth-, and fourth-generation gaps. These excitations are of Boltzmann
nature and govern the specific heat until the temperature becomes of
the order of $T_F$.  At this point electrons are able to jump up over
the third-generation gap between the eighth and ninth bands.  The
high-$T$ ($T > 3^{-4}$) oscillations are associated to electronic
excitations through the third-, second-, and first-generation gaps.  Similar
considerations apply to the complementary case $N^*=57/64$.  The
horizontal lines in Fig.~8 correspond to our predictions for the
average specific heat, which take into account the effective number of
electrons or holes that contribute to the specific heat in each
regime.  (The fact that the case $N^*=7/64$ presents one less
oscillation than its complementary is due to an overlap of 
temperature scales in the transition from hole to electron behavior.)
\begin{figure}[ht]
\hspace{0.0pc} 
\epsfxsize=7.0cm
\epsfbox[77 193 482 555]{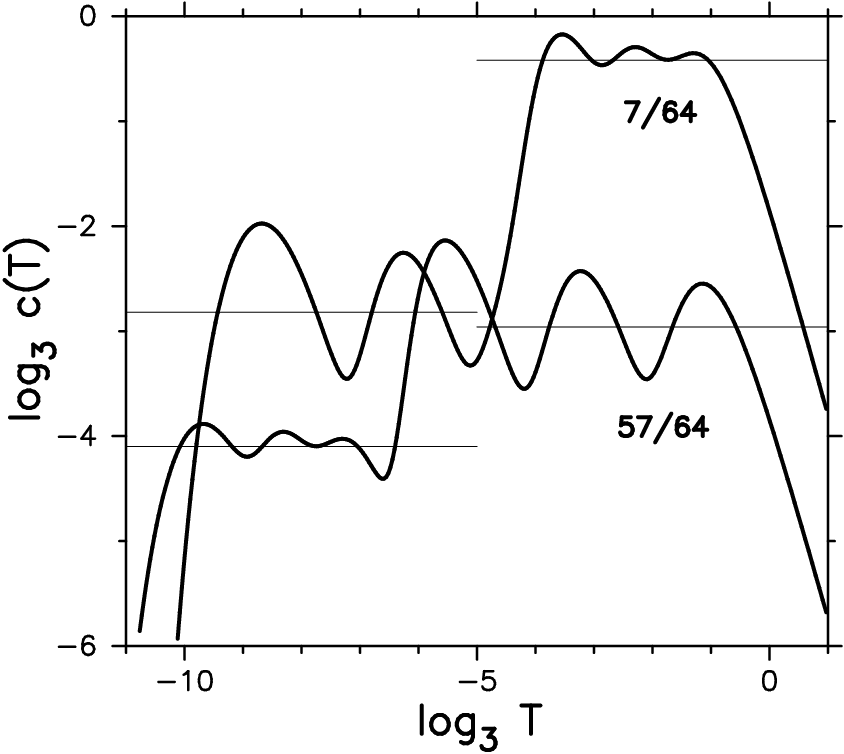}
\vspace*{1.0pc}
\caption{Analogous to Fig.~7 but for the particle 
numbers $N^*=7/64,57/64$ (see text).}
\end{figure}
The general case of an arbitrary number of particles shows a
non-trivial mixture of the simple behaviors described before.  However,
the gross features of the specific heat for arbitrary $N$ can be
qualitatively (and sometimes quantitatively) understood as being a
reflection of sequences of electronic and hole excitations that
alternate themselves in producing the oscillatory patterns shown in
Fig.~9.
\begin{figure}[ht]
\hspace{0.0pc} 
\epsfxsize=7.0cm
\epsfbox[77 193 482 555]{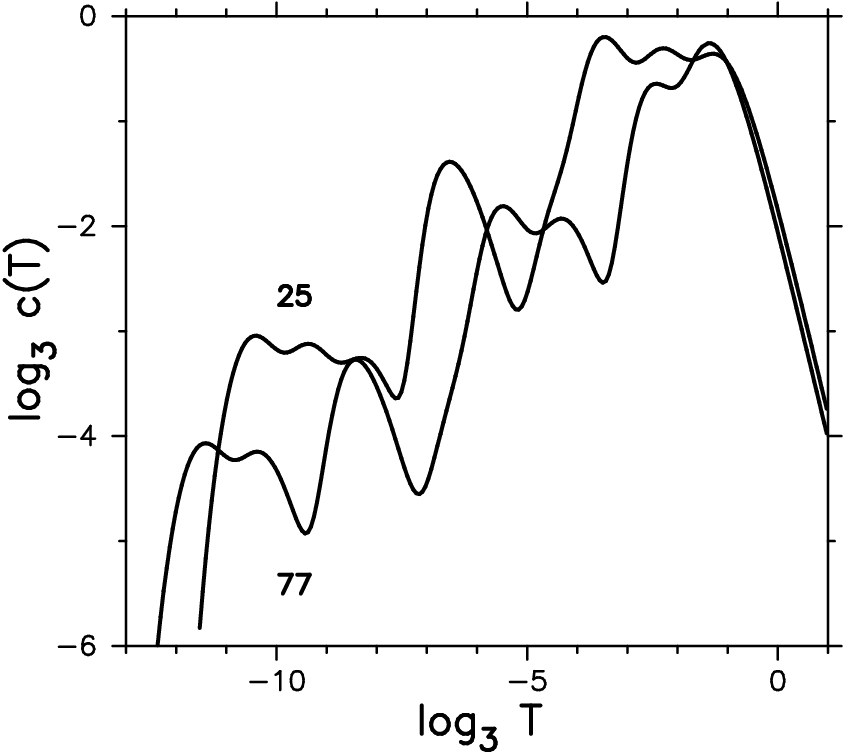}
\vspace*{1.0pc}
\caption{Fermions in a fractal spectrum. Log-log plot of the specific
heat versus temperature for the numbers $N=25,77$, and
the scale factors $r_1=1/3$, $r_2=1/9$ ($n=7$).}
\end{figure}
Formulations based on smooth approximations to the density of states,
analogous to those made for bosons, are bound to fail in the fermionic
case. The intrinsic discontinuity of the Fermi problem, which is
critically enhanced by the dual scaling respect to the lower and upper
edges of the gaps, is in essence non compatible with smooth
approximations (with the exception of too special cases).


\section{Concluding remarks}


We have analyzed the quantum statistics of model systems exhibiting
two-scale fractal spectra, with special emphasis on the structure of
the specific heat. Our findings extend previous results on classical
statistics to show that the thermodynamical manifestations of spectral
fractality are robust with respect to the inclusion of quantum
symmetries. The general scenario for Boltzmann and bosonic particles
can be summarized as follows. In spite of the very fragmented structure
of the real density of states, a formulation which starts from a smooth
approximation but takes into account the coarsest log-periodic
fluctuations is sufficient for a good description of the specific heat
and other averaged quantities. In the case of fermions, even if a
treatment which is uniform in the particle number is not possible, many
features of the problem can be understood with the help of the simple
results for the Boltzmann case.

Our analysis was limited to two-scale fractals. Previous
results\cite{vallejos98} indicate that provided the scaling towards the
inferior limit of the spectrum is uniform, the inclusion of additional
scales will not change our conclusions in what concerns the bosonic
case. Fermions, however, can feel the scaling with respect to any point
of the spectrum. So, the complexity of the fermionic problem would be
proportional to the number of relevant scales, in spite of uniform
zero-energy scaling.

Let us finish by mentioning that the log-periodicities described in
this paper may be observable in real physical systems, e.g. in the
Fibonacci superlattices\cite{merlin85,todd86}. In fact, although our
discussion was restricted to perfect deterministic systems, both
experiments and numerical simulations by Todd et al.\cite{todd86}
indicate that the hierarchical organization of the electronic bands may
be preserved even if substantial amounts of (random) disorder are added
to the system. From a more general point of view, Saleur and
Sornette\cite{saleur96} have demonstrated that the connection between
discrete scale invariance and log-periodic oscillations is robust
with respect to the presence of disorder.


\acknowledgements

We acknowledge Brazilian agencies FAPERJ and PRONEX for financial
support. Also acknowledged is the kind hospitality at the Centro Brasileiro
de Pesquisas F\'{\i}sicas, where part of this work was done.



\end{multicols}

\end{document}